DOI:

# MATHEMATICAL MODELLING OF ASTROPHYSICAL OBJECTS AND PROCESSES

**Ivan L. Andronov, Vitalii V. Breus, Larysa S. Kudashkina**

> "The book of Nature is written
> in the language of Mathematics"
> *Galileo Galilei (15.2.1564 – 8.1.1642)*

## INTRODUCTION

Mathematical methods are the base of each branch of science. Similar equations may be used for a wide variety of processes.

At the Department "Mathematics, Physics and Astronomy", there are some scientific directions of research, which may be generally described as "Mathematical modeling of processes and systems". The International Astronomical Union[1] listed numerical directions of impact of astronomy into studies and technical applications to the society. Among them there are medicine, climate change, computing, time keeping, imaging, communication, Wi-Fi. One may add (planar, spherical, multi-dimensional) geometry, coordinate keeping, mathematical modeling, high-energy physics, space research.

Currently, there are huge international projects, which are also present in Ukraine – "Virtual Observatory"[2], "AstroInformatics"[3]. These topics are related to the international project ILA ("Inter-Longitude Astronomy")[4], which consists of a series of smaller projects on concrete variable stars of different types, binary systems with possible exo-planets or stellar-mass campaigns or pulsators.

The corresponding scientific school of researchers of variable stars[5,6] was founded by Prof. Vladimir Platonovich Tsesevich (11.10.1907-28.10.1983), the famous Scientist, Lectures, Popularizer and Organizer of Science in our country.

---

[1] Downer B., Burton M., van Dishoeck E., Russo pp. (2019). From Medicine to Wi-Fi: Technical Applications of Astronomy to Society, Published by the IAU, https://www.iau.org/static/archives/announcements/pdf/ann19022a.pdf
[2] Vavilova, I. B. Pakuliak L. K., Protsyuk Yu. I. et al. (2011). Ukrainskaya virtual'naya observatoriya (UkrVO). Sovremennoe sostoyanie i perspektivy razvitija ob`edinennogo arhiva nabljudenij [Ukrainian Virtual Observatory: Current Status and Perspectives of Development of Joint Archives of Observations]. *Kosmicheskaya nauka i tehnologiya*: vol. 17, no. 3, pp. 74-91. ADS: 2011KosNT..17d..74V
[3] Vavilova I. B., Yatskiv Ya. S., Pakuliak L.K. (2017). UkrVO Astroinformatics Software and Web-services. *Astroinformatics, Proceedings of the International Astronomical Union, IAU Symposium*, Vol. 325, pp. 361-366, DOI: https://doi.org/10.1017/S1743921317001661 . ADS: 2017IAUS..325..361V
[4] Andronov I. L., Andrych K. D., Antoniuk, K. A. et al. (2017). Instabilities in Interacting Binary Stars. *Astron. Soc. Pacif. Conf. Ser.,* Vol. 511, pp. 43-50. ADS: 2017ASPC..511...43A
[5] Andronov I.L. (2017). Odessa Scientific School of Researchers of Variable Stars: from V.P.Tsesevich (1907-1983) to Our Days. *Odessa Astronomical Publications*, vol. 30, pp.252-255. ADS: 2017OAP....30..252A

In this review, we will highlight the aspects of the Astroinformatics related to:
- Statistically optimal data analysis – elaboration of improved algorithms and programs with a complete set of formulae;
- Highlights of the monitoring and analysis of binary systems;
- Highlights of the monitoring and analysis of cataclysmic variables;
- Highlights of the monitoring and analysis of pulsating variables;

**TIME SERIES ANALYSIS: BEYOND THE OVERSIMPLIFIED METHODS**

**General equations.** There are many methods for the time series analysis (generally, the data analysis), which are described in many hundreds (or even thousands) of monographs and textbooks (see the classical ones, e.g. [7,8,9,10,11]). There is no unique solution for many types of variability, especially, in a case of few simultaneously acting physical mechanisms. There are special methods, which are oriented to separate types of data. Do our usual advice is to use a net of different methods to check, if they will produce self-consistent results.

Numerous electronic tables provide a wide variety of functions for making graphs and stastistical studies of the data. There is also a possibility to show approximations using the function from some list. The user may choose a desired function and visually estimate a quality of the approximation. However, the choice is based on "estetic" feeling, or, maximally, on the value of $r^2$, the square of the correlation coefficient between the original data and the approximated values. The approximations do not take into account possible weights (different accuracy) of the data.

Generally, one may write an approximation as a function of the data and the parameters (coefficients)

$$x_c(t) = \sum_{\alpha=1}^{m} C_\alpha \cdot f_\alpha(t; C_\gamma), \qquad (1)$$

where $\gamma = m + 1, \ldots, L$. The coefficients may be formally classified as "linear" $(1, \ldots, m)$ and "non-linear" $(m + 1, \ldots, L)$ ones. So $L = m + p$ is the total number of coefficients, and $p -$ is the number of "non-linear coefficients". We number them in such a way, that two types will be separated into two groups.

The same approximation may be (typically) written with a different number of linear and non-linear coefficients, e.g., for a polynomial

$$x_C(t) = C_1 + C_2 \cdot t + C_3 \cdot t^2 + \cdots + C_m \cdot t^{m-1} =$$
$$= \tilde{C}_1 + \tilde{C}_2 \cdot (t - \tilde{C}_m)^2 + \cdots + \tilde{C}_{m-1} \cdot (t - \tilde{C}_m)^{m-1}. \quad (2)$$

The second type has only one ($p = 1$) non-linear coefficient $\tilde{C}_m$, which corresponds to an extremum or a stationary point, as $\frac{\partial x_c(t; C_\gamma)}{\partial t} = 0$.

Although the basic functions in Eq. (1) are defined for $\alpha = 1, \ldots, m$, this definition may be extended to a fill range ($\alpha = 1, \ldots, L$):

$$f_\alpha(t; C_\alpha) = \frac{\partial x_C(t; C_\gamma)}{\partial C_\alpha}. \quad (3)$$

The coefficients should be determined to minimize the generalized squared distance between the observations $x_k$ and the approximation $x_{Ck} = x_c(t_k; C_\gamma)$.

The test function currently may be written in a general case as[12]

$$\Phi_L = \sum_{kj=1}^{n} p_{kj} w_{kj} \varepsilon_k \varepsilon_j. \quad (4)$$

Here $w_{kj} = \sigma_0^2 \mu_{kj}^{-1}$ is the matrix proportional to the inverse of the covariation matrix $\mu_{kj}$ of errors of measurements, $p_{kj} = p(t_k, t_j, t_0, \Delta t)$ is the additional weight function, which is typically used in the wavelet analysis and in the dynamical spectra. The deviations of the "observed" ($O$) data from the "calculated" ($C$) ones are $\varepsilon_k = x_k - x_{Ck}$. There is also a relation $\mu_{kj} = \sigma_0^2 w_{kj}^{-1}$. Contrary to usual simplified matrix equations, in the advanced formulae, we have to write indices for clarity.

The coefficients $C_\alpha$ should minimize the test function $\Phi_L$, so $\frac{\partial \Phi_L}{\partial C_\alpha} = 0$, and the linear coefficients may be determined as

$$C_\alpha = \sum_{\beta=1}^{m} A_{\alpha\beta}^{-1} B_\beta, \quad (5)$$

$$A_{\alpha\beta} = \sum_{kj=1}^{n} p_{kj} w_{kj} f_\alpha(t_k) f_\beta(t_j), \quad (6)$$

$$B_\beta = \sum_{kj=1}^{n} p_{kj} w_{kj} x_k f_\beta(t_j), \tag{7}$$

**Error estimates of the parameters and functions.**

The covariation matrix of the errors of the coefficients is

$$R_{\alpha\beta} = \sum_{\gamma\varepsilon=1}^{L} A_{\alpha\gamma}^{-1} A_{\varepsilon\beta}^{-1} \sum_{kj=1}^{n} p_{kj}^2 w_{kj}^2 f_\gamma(t_k) f_\varepsilon(t_j), \tag{8}$$

Every function $G(C_\alpha)$ has a mathematical expectation of the variance

$$\sigma^2[G(C_\alpha)] = \sum_{\alpha\beta=1}^{L} R_{\alpha\beta} \cdot \frac{\partial G}{\partial C_\alpha} \cdot \frac{\partial G}{\partial C_\beta}. \tag{9}$$

Particularly, $\sigma^2[C_\alpha] = R_{\alpha\alpha}$, so the accuracy of the coefficient is $\sigma[C_\alpha] = \sqrt{R_{\alpha\alpha}}$. Similarly, the variance of the derivative of the order $s$ of the smoothing function

$$x_C^{(s)}(t) = \frac{\partial^s x_C(t)}{\partial t^s} = \sum_{\alpha=1}^{m} C_\alpha \cdot f_\alpha^{(s)}(t; C_\gamma), \tag{10}$$

$$\sigma^2\left[x_C^{(s)}(t)\right] = \sum_{\alpha\beta=1}^{L} R_{\alpha\beta} \cdot f_\alpha^{(s)}(t; C_\gamma) \cdot f_\beta^{(s)}(t; C_\gamma). \tag{11}$$

The matrices are symmetrical: $\mu_{kj} = \mu_{jk}$, $w_{kj} = w_{jk}$, $p_{kj} = p_{jk}$, $A_{\alpha\beta} = A_{\beta\alpha}$, $R_{\alpha\beta} = R_{\beta\alpha}$. The upper limit in these sums is $m$ for the approximation and its derivatives, and $L = m + p$, if the initial values of "non-linear" parameters are determined using the method of differential corrections (sometimes, called the "Newton-Raphson method").

It should be noted that the common style is to publish the values of the accuracy (statistical error estimates) of the coefficients $\sigma[C_\alpha] = \sqrt{R_{\alpha\alpha}}$. From these values, it is not possible to restore non-diagonal elements $R_{\alpha\beta}$, i.e for $\alpha \neq \beta$. In the absence of this information, these values are *suggested* to be zero. In other words, the correct matrix $R_{\alpha\beta}$ is *formally* replaced by $\delta_{\alpha\beta} R_{\alpha\beta}$. This may lead to wrong results as compared with correct ones in Eq. (9, 11).

**Differential corrections**

According to the Marquard's[13,] modification of the Levenberg's[14,] method, the differential corrections may be computed as

$$\Delta C_\alpha = \sum_{\beta=1}^{L} (A_{\alpha\beta} + \lambda \cdot \delta_{\alpha\beta} \cdot A_{\alpha\alpha})^{-1} \cdot \Delta B_\beta, \tag{12}$$

$$\Delta B_\beta = \sum_{kj=1}^{n} p_{kj} w_{kj} \varepsilon_k f_\beta(t_j), \tag{13}$$

where and $\delta_{\alpha\beta}$ − is the Kronecker symbol. At each iteration, the values of non-linear parameters are replaced in a sense $C_\alpha \coloneqq C_\alpha + \Delta C_\alpha$. The regularization parameter is $\lambda \geq 0$. Typically, $\lambda = 0$ for non-degenerate matrix of normal equations with components $A_{\alpha\beta}$, thus $(A_{\alpha\beta} + \lambda \cdot \delta_{\alpha\beta} \cdot A_{\alpha\alpha})^{-1}$ in Eq. (12) may be replaced with $A_{\alpha\beta}^{-1}$ as in Eq. (5) [15]. For the case of nearly degenerate matrix, one may choose $\lambda \gg 1$. This may increase a number of iterations needed to reach the desired accuracy. In any case, for correct determination of the matrix $R_{\alpha\beta}$ (Eq. (5)), the parameter $\lambda$ is not used[16].

---

[13] Marquardt, Donald (1963). An Algorithm for Least-Squares Estimation of Nonlinear Parameters. *SIAM Journal on Applied Mathematics*. 11 (2): 431–441. DOI: https://doi.org/10.1137/0111030
[14] Levenberg K. (1944). A Method for the Solution of Certain Non-Linear Problems in Least Squares. *Quarterly of Applied Mathematics*. 2 (2): 164–168. DOI: https://doi.org/10.1090/qam/10666

### Simplification

Only under two assumptions, this matrix may be simplified to

$$R_{\alpha\beta} = \sigma_{0L}^2 A_{\alpha\beta}^{-1}, \tag{14}$$

$$\sigma_{0L}^2 = \frac{\Phi_L}{n-L}. \tag{15}$$

The first assumption is $p_{kj} = \text{const} > 0$ for all data ("global" approximation) or for a subset of data ("local" approximation, "rectangular filter"). The second assumption is that the mathematical expectation $w_{kj} = \sigma_0^2 \mu_{kj}^{-1}$. As the matrix $\mu_{kj}$ is typically unknown, especially the non-diagonal elements, which correspond to non-zero correlations between the deviations of the observations, there are simplifications. The reasonable simplification is that these matrices are diagonal, i.e.

$$\mu_{kj} = \delta_{kj} \cdot \mu_{kk} = \delta_{kj} \cdot \sigma_k^2, \tag{16}$$

where $\sigma_k = \sigma[x_k]$ − is the accuracy of the measurement $x_k$. Then

$$w_{kj} = \delta_{kj} \cdot w_{kk} = \delta_{kj} \cdot \frac{\sigma_0^2}{\sigma_k^2} \tag{17}$$

Then the sum becomes over only one index:

$$G = \frac{\Phi}{\sigma_0^2} = \sum_{k=1}^{n} \left(\frac{x_k - x_{Ck}}{\sigma_k}\right)^2 = \sum_{k=1}^{n} w_{kk} \cdot \varepsilon_k^2. \tag{18}$$

Next often assumption is that the distribution of errors of measurements is the $n-$ dimensional normal distribution, and $\varepsilon_k/\sigma_k$ are random normal values with a zero

---

[15] Andronov I.L. (1994) (Multi-) Frequency Variations of Stars. Some Methods and Results. *Odessa Astronomical Publications*, vol. 7, pp. 49-54. ADS: 1994OAP.....7...49A

[16] Andronov I.L. (2003) Multi-periodic versus noise variations: mathematical methods. *Astronomical Society of the Pacific Conf. Ser.* Vol. 292, pp. 391-400. ADS: 2003ASPC..292..391A

mathematical expectation and unit variance. In this case, the random value $G$ is expected to obey the $\chi^2_{n-L}$ distribution[17].

**Oversimplification**

Assuming that the weights $w_{kk}$ of the observations are equal, one may set them to unity, so the matrix becomes an unit matrix: $w_{kj} = \delta_{kj}$, and thus may be omitted in the equations above.

This "unweighted" oversimplified model is realized in the electronic tables like Microsoft Excel, Libre/Openoffice Calc, Kingsoft Spreadsheets, GNU Gnumeric et al. There are some polynomial, exponential, power approximations.

Unfortunately, these programs do not yet allow estimation of the errors of the coefficients $C_\alpha$ in the smoothing function $x_C$ and its derivatives. This may be done using a program code (macros, script) in some computer language, including the VBA present in Excel.

It should be noted that it is convenient to use matrix designations in theoretical work. However, for programming, it is more suitable to compute a trial vector of values of basic functions for a current parameter.

**"Statistically optimal" number of parameters $m$**

The assumption of the $n$ − dimensional distribution of errors of measurements allows to determine False Alarm Probability ($FAP$) that the variation of the approximation may not be resulted by random errors. The less ($FAP$) is, the more statistically significant is. So the "statistically optimal" number of parameters $m$ in a case of choosing the model with different number of (unknown) parameters $m$.

The "quality of approximation" for $m$ and $m − q$ parameters may be compared using the test function from Eq. (4):

$$F = \frac{n-m}{q} \cdot \left(\frac{\Phi_{m-q}}{\Phi_m} - 1\right) \qquad (19)$$

For a common approach of a normal distribution of errors of measurements, the value of $F$ has the Fisher[18] probability distribution function with $q$ and $n − m$ degrees of freedom. One may choose some limiting False Alarm Probability ($FAP$), and then the number of parameters $m$ will correspond to the maximal value, when still $FAP < FAP_{\text{critical}}$.

The same FAP estimate may be obtained using

$$B = \left(1 - \frac{\Phi_m}{\Phi_{m-q}}\right) = \frac{q \cdot F}{(n-m) + q \cdot F}. \qquad (20)$$

Similarly,

---

[17] Forbes C., Evans M., Hastings N., Peacock B. *Statistical Distributions*. 4th ed. — Wiley, 2011. — 231 pp.
[18] Fisher R.A. (1954). *Statistical Methods for Research Workers*, 12th Ed. Oliver and Boyd, Edinburgh, 356 pp.

$$F = \frac{(n-m) \cdot B}{q \cdot (1-B)} \qquad (21).$$

For a pure noise, the value $B$ obeys the $B$ (Beta) probability distribution function.

Although in these equations, there are no limitations to the value of $q$, in practice, $q = 2$ for trigonometric polynomials. Numerous publications on a topic may be found for "ANOVA" (=ANalysis Of VAriances).

Next criterion may be used is the minimization of the accuracy estimate of the approximation $x_C(t_0)$ (or its derivative of any order) at some fixed point $t_0$ using Eq. (11). This may be reasonable, e.g., if to make a forecast at some point.

For the approximation of all data, one may use the estimate of the accuracy $\sigma\left[x_{CL}^{(s)}\right]$ of the approximation or its derivative of order $s$ using the r.m.s. value:

$$\sigma^2\left[x_{CL}^{(s)}\right] = \frac{1}{n} \sum_{k=1}^{n} \sigma^2\left[x_C^{(s)}(t_k)\right]. \qquad (21)$$

In the simplified model, even with a generally non-diagonal matrix $w_{kj}$,

$$\sigma^2\left[x_{CL}^{(s)}\right] = \frac{L}{n} \sigma_{0L}^2 \qquad (22)$$

Often, for small $L$, the value of $\sigma_{0L}^2$ is decreasing with decreasing systematical deviations of the data from the approximation. Then, for good approximations, $\sigma_{0L}^2$ shows a "standstill" (with fluctuations due to sample effects), so $\sigma^2\left[x_{CL}^{(s)}\right]$ has a minimum at some $L$, which corresponds to the "statistically optimal approximation", according to this criterion. Often, the "statistically optimal" value of $L$ is less than that obtained according to the FAP

Other modification may be for the r.m.s. accuracy at some interval from $t_{start}$ to $t_{end}$,

$$\sigma^2\left[x_{CLI}^{(s)}\right] = \frac{1}{t_{end} - t_{start}} \int_{t_{start}}^{t_{end}} \sigma^2\left[x_C^{(s)}(t)\right] dt. \qquad (23)$$

As an example, this may be recommended for approximations of data with large gaps or large inhomogeneity of distribution of times. For periodic functions, it may be recommended to determine r.m.s. value over a complete period[15].

Sometimes the criterion based on discrete or continuous averaging of the error estimates leads to very small $L$. Typically, if the the signal is very noisy. Formally, the best accuracy may correspond to $L = 1$, e.g. the weighted mean value. No variations are present at the approximation in this case. If looking for variability, we propose to use the "amplitude-based" SNR (S/N= signal-to-noise ratio). It may be defined as the ratio of the weighted r.m.s. deviation of the data from the mean value to the r.m.s. accuracy of the approximation $\sigma[x_{CL}]$.

The weighted r.m.s. deviation of the data from the approximation $\sigma[x_{CW}]$ is

$$\sigma^2[x_{CW}] = \frac{\sum_{k=1}^{n} w_{kk} \cdot \varepsilon_k^2}{\sum_{k=1}^{n} w_{kk}}, \quad (24)$$

and may be generalized similarly to Eq. (4).

**Weight function in wavelet analysis-like methods**

The weight (or "window") function $p_{kj} = p(t_k, t_j, t_0, \Delta t)$ is written in a general form. The dependence on both times $t_k, t_j$ is typically neglected, so the matrix is diagonal: $p_{kj} = \delta_{kj} p_{kk}$. Often, it is suitable to decrease the number of parameters to a single function $p_{kk} = p(z_k)$ of a dimensionless parameter

$$z_k = \frac{t_k - t_0}{\Delta t}. \quad (25)$$

In the "wavelet terminology", $t_0$ may be called as a "shift", and $\Delta t$ — "scale"[19].

In other words, $t_0$ corresponds to a "trial" time, which "scans" the interval with a characteristic "half-width" $\Delta t$.

In practice, the most common shape of the function is a "constant" ($p_{kj} = 1$) one corresponding to "global" approximations.

Next popular one is the "rectangular" window $p(z) = 1$ only for $-1 \leq z \leq +1$, i.e. for the data inside the interval $[t_0 - \Delta t, t_0 + \Delta t]$, neglecting the data outside ($p(z) = 0$). This is used in the "running mean" (also called "moving average") smoothing of data, as well as for numerous digital filters applied to regularly spaced data, for which $t_k - t_j = (k - j) \cdot \delta$. The classical monograph on "Digital Filters" by Hamming[20], was followed by dozens of other books on linear[21] and non-linear[22] filtering.

The statistically accurate set of expressions for a general case of irregularly spaced data (and any basic and weight functions) was presented by Andronov[23].

The local approximation $x_C(t, t_0, \Delta t)$ is computed for different $t_0$, the parameters $C_\alpha$ are computed using a complete form of the method of the Least Squares (LSQ), but only one value is used at $t = t_0$:

$$\tilde{x}_C(t_0, \Delta t) = x_C(t_0, t_0, \Delta t). \quad (26)$$

Generally, these functions cross at $t = t_0$, so the derivatives of are not equal:

---

$$\left.\frac{\partial \tilde{x}_C(t_0,\Delta t)}{\partial t_0} \neq \frac{\partial x_C(t,t_0,\Delta t)}{\partial t}\right]_{t=t_0}. \tag{27}$$

In the "running mean", the function $p(z)$ is rectangular. So, every data point, which comes inside (or outside) the interval, will change the approximation abruptly, and $\tilde{x}_C(t_0, \Delta t)$ is discontinuous. This is neglected in the classical "running mean", as the grid of $\tilde{x}_C(t_0, \Delta t)$ exactly coincides with the times $t_k$.

Thus it was proposed to use a function with zero value and its derivative at the ends of the interval: $p(\pm 1) = 0$, $p'(\pm 1) = 0$. There may be an infinite number of non-negative functions with additional properties. One of the simplest ones is $p(z) = (1 - z^2)^2$ (being zero for $|z| > 1$).

Combined to basic functions $1, z, z^2$, this makes a parabolic local approximation $x_C(t_0, t_0, \Delta t)$, and a smooth function $\tilde{x}_C(t_0, \Delta t)$. Thus the method is called the "running parabola". It allows fitting of the third order polynomial exactly at a grid of regular points, when $t_k$ are located symmetrically in respect to $t_0$ within the interval of $p(z) > 0$.

The only free parameter defining the quality of the approximation is $\Delta t$. For small $\Delta t$, the approximation has large-amplitude apparent waves, which are not statistically significant due to large error estimates of the approximation. This is similar to very large number of parameters $L$. For large $\Delta t$, the approximation asymptotically tends to a "global" one with constant $p(z) = 1$ and shows minimal variability. As in the electronic tables for the degree of the polynomial, one may choose $\Delta t$ "esthetically", by visual comparison of the data with the approximation.

**Criteria for Statistically Optimal Determination of $\Delta t$**

Similarly to determination of statistically optimal number of parameters discussed earlier, there may be few basic functions describing the quality of the fit, which may be rewritten as:
- $\sigma_0(\Delta t)$ – the unbiased weighted r.m.s. estimate of the deviations of the data $x_k$ from the approximation $x_C(t_k, t_k, \Delta t)$ at this time;
- $\sigma_{xC}(\Delta t)$ – the r.m.s. accuracy estimate of the approximation at times $t_k$;
- SNR=S/N – the amplitude signal-to-noise ratio.

The detailed expressions are presented in [23]. One has to compute a "scalegram" for a set of values. We recommend $\Delta t = \Delta t_{\min} \cdot 10^{i/20}$, $i = 0, \ldots, i_{\max}$. This may be done in the computer program OO[24], or in the more recent software MAVKA[25] with limits of $\Delta t$ by (data related automatic) default, or setting by an user.

There is no analogue of the Fisher's criterion, as there is no integer parameter determining the quality of the observations. Thus one may recommend to use $\Delta t$ corresponding to maximum of the SNR (preferred), or to minimum of $\sigma_{xC}(\Delta t)$ (which typically corresponds to larger value of $\Delta t$, than that from the SNR). The dependence $\sigma_0(\Delta t)$ is typically negligible for $\Delta t \ll P$ (where $P$ — is the period of a test signal), and then has an increase to a peak, and then a slight decrease to another standstill at $\Delta t \gg P$. By fitting the theoretical dependence to the sample function, one may determine the characteristic timescale (effective period), and an effective semi-amplitude of the variations. As a characteristic of stability of the period, the "characteristic width" of the scalegram may be introduced. This was done for 173 semi-regular pulsating variable stars[26].

For the fractal-type variability, the dependence obeys the power law[27]: $\sigma_0(\Delta t) \propto (\Delta t)^\gamma$. For the determination of the characteristic time scale (cycle length) and semi-amplitude of quasi-periodic oscillations (QPO), Andronov[16] proposed a "$\Lambda$ − scalegram" analysis based on the derivative $\Lambda(\Delta t) = d\sigma_0^2(\Delta t)/d\Delta t$.

For a single value at some point, the criteria may be replaced by least error estimate of the approximation.

**"Running Sine"**

This running approximation is based on the function
$$x_C(t, t_0, \Delta t) = a - R \cos(2\pi(t - T_0)/P) \qquad (28)$$
with 3 parameters, which are functions of either $t_0$, or $\Delta t$: $a$ − the average value of the approximation over the period $P$, $R$ − the semi-amplitude (in some literature, also called "amplitude") and $T_0$ − moment of minimum of the approximation (chosen to be nearest to $t_0$). Detailed expressions and numerical examples are presented in the review[28].

In this method, the weight function $p(z) = 1$ for $|z| \leq 1$. It should be recommended to choose $\Delta t = iP/2$, where $i = 1, 2$, or (with very large gaps in the observations) 4, 8, 16… This will cease the influence of harmonics of periodic variability. The smaller $i$ is, the larger are error estimates of these parameters.

The method is effective for analysis of signals with slow modulation of each of these parameters. Moreover, if the period is variable or different from the initial value $P$, one may analyze, using the dependence of phase

---

$$\phi = (T_0 - T_{00})/P - \text{int}((T_0 - T_{00})/P + 0.5) \qquad (29)$$

on trial time $t_0$. Here $T_{00}$ — is some initial value of the initial epoch.

This method may be improved for a running trigonometric polynomial:

$$x_C(t, t_0, \Delta t) = a - \sum_{j=1}^{s} R_j \cdot \cos(2\pi j \cdot (t - T_{0j})/P) \qquad (30)$$

In some cases of symmetrical signals, one may decrease a number of unknowns by fixing $s$ parameters $T_{0j}$ to one unknown parameter $T_{00}$. It may be computed, for each trial $t_0$, using differential corrections, as described above.

**Morlet –type wavelet**

The "true" Morlet wavelet was proposed for infinite continuous signals. For irregularly spaced discrete signals, the integrals were replaced by corresponding sums[29]. This is a common method for evenly sampled data, and is included in some software.

The problems arise for unevenly sampled data, for which the noise at the wavelet map is large even for a pure cosine signal. This noise may be reduced using the method of the least squares[30]. This running approximation is based on the function (28) in a form

$$x_C(t, t_0, \Delta t) = C_1 + C_2 \cos(z) + C_3 \sin(z) \qquad (31)$$

and a Gaussian weight function $p(z) = \exp(-c \cdot z^2)$. Here $z = \omega \cdot (t - t_0)$, $\omega = \frac{2\pi}{P}$. A free parameter $c$ is typically set theoretically to $\frac{1}{8\pi^2}$, or practically to a close value $\frac{1}{80} = 0.0125$.

Andronov[31,32] introduced a complementary test function S for the wavelet map, the peak of which is not shifted in a period, contrary to 3 functions (R, WWT, WWZ) used by [30], a wavelet approximation and wavelet periodograms. The use of the least squares instead of oversimplified sums allowed decreasing of the noise by a factor of few times to (even) few dozen times.

Depending on stability of the signal variations and the noise of measurements, the value of $c$ may be changed[33,34]. Asymptotically, for $c \to 0$, the approximation

---

becames a "global" one, thus no modulations of the parameters may be seen. Alternately, for $c \to \infty$, the approximation becomes the same, as of the "running parabola". Thus, the wavelet analysis is "intermediate" between the cases of "global" and "narrow local" approximations.

Taking into account numerous seasonal gaps in ground-based astronomical observations, the advantages of using the Gauss function as the weight function vanish. It may be recommended to use a wavelet with a compact weight function, like that used in "running parabola" $p(z) = (1 - z^2)^2$ (being zero for $|z| > 1$)[35], or even that closer to a rectangular shape: $p(z) = (1 - z^4)^2$ (or even larger power).

**Periodogram analysis**

The methods are based on computation of many phase curves for a set of trial periods $P$ (or frequencies $f = 1/P$) and determination of the one corresponding to the "best" curve. For each trial $f$, the phases are computed as

$$\phi = (t - T_0)/P - \text{int}((t - T_0)/P). \tag{32}$$

Contrary to (29), in this definition, $0 \leq \phi < 1$, and the initial epoch $T_0$ may be chosen arbitrarily (e.g. the beginning of the measurements $t_1$). Obviously, the data are repeated with a period 1, in this scale, so, formally, e.g. $\phi = -2.3, 123.7$ correspond to the same phase $\phi = +0.7$ "in the main interval" from 0 to 1.

Then the methods may be split into two groups" the "point-point" (or "non-parametrical") ones (see a comparative review of most popular methods)[36], and "point-curve" ("parametrical") ones[15,16,12]. The first method was implemented in VSCalc[37,38].

Some methods are included in the software MCV[39,40], particularly, the unique algorithm of the periodogram analysis, where multi-harmonic variations superimposed onto a trend (polynomial of order $s_0$) are taken into account:

$$x_C(t) = C_1 + \sum_{\alpha=1}^{s_0} C_{\alpha+1} \cdot \tilde{t}^\alpha + \sum_{\alpha=1}^{s} \left( C_{2\alpha+s_0} \cdot \cos(z) + C_{2\alpha+s_0+1} \cdot \sin(z) \right), \tag{33}$$

where $z = \omega \cdot \tilde{t}$, $\tilde{t} = t - t_{\text{mean}}$.

We recommend to subtract a sample mean time $t_{\text{mean}}$ to decrease the degeneracy characteristic of the matrix of normal equations for further differential corrections, when $\omega$ is corrected using differential corrections and thus becomes a "non-linear" parameter $C_{1+s_0+s}$.

The test function used for this type of the periodogram analysis is similar to that in Eq. (20):

$$S(f) = 1 - \frac{\Phi_{1+s_0+2s}}{\Phi_{1+s_0}} \tag{34}$$

For a pure Gaussian noise, it has a Beta distribution with the numbers of degrees of freedom $2s$ and $(n - 1 - s_0 + 2s)$. The statistically optimal value of frequency $f$ corresponds to the highest maximum of $S(f)$.

It is important to note, that the coefficients describing the trend, are also functions of frequency. This is significantly different from the "detrending", when the trend is removed from the data before the periodogram analysis. This is especially important, if the duration of measurements is comparable to the period of the signal, e.g. the superhumps in the SU UMa-type stars.

The fastest periodogram analysis is based on the cosine approximation without any trend, so $s_0 = 0$, $s = 1$. However, for signals with highly a-sinusoidal shape, one has to increase $s$. The algorithm and program making it (with correction of the frequency), was described in[15]. For the fixed $s$, the correction may be done in the MCV.

Also it is possible to apply a model of multi-harmonic variations with 1, 2 or 3 periods, also generally superimposed onto a polynomial trend. The accuracy of the coefficients and the approximation are available in the output files.

In this software, there is also a possibility are many methods oriented on data analysis of variable stars. However, it may be used also at minimum efficiency – as the default viewer of (multi-column) data files with automatic scaling.

**Special shapes (patterns, templates, profiles)**

Even for periodic signals, the shape may be very a-sinusodal, with intervals of significant changes. This is a typical situation, e.g., for the Algol-type stars, which show distinct eclipses in an addition to low-amplitude smooth variations outside eclipses. The interval of data then should be split into few parts.

Often, the interval is split, if using polynomial splines and their improvements[41]. This method allowed determining best periods and approximations of the light curve, as well as to check for possible period changes.

---

[41] Andronov I.L. (1987). Smoothing the "smoothing" cubic spline functions. – *Publications of the Astronomical Institute of the Czechoslovak Academy of Sciences*, No. 70, pp. 161 – 164. ADS: 1987PAICz..70..161A

The concept of polynomial splines with alternating order was proposed by Andronov[16], with sequences of 2-0-2-0 for eclipsing binary stars (with two eclipses) and 2-3 for pulsating RR Lyr-type stars (with sharp rise and slow fall). This was applied for automatic classification of 863 newly discovered variable stars from the Hipparcos-Tycho space observations[42].

The improved accuracy of observations for new variable stars needed elaboration of models with better quality of approximation. This leads to increase of the number of parameters $m$, which, obviously, should be smaller than that for the trigonometric polynomial. Andronov[43] proposed the "New Algol Variable" (NAV) algorithm, which combined the trigonometric polynomial of second order (which fits the possible effects of ellipticity, reflection and spots) and "shapes" of eclipses. For a given initial epoch and a period, there are $m = 7$ "linear" parameters and 3 "non-linear" parameters, but the initial epoch and a period may be added. A total number of parameters $L = 12$ corresponds to a trigonometric polynomial of degree $s = 5$, which is rather good for the smooth curves of EW and EB=type variables. For EA-type (Algols), the number of parameters for a trigonometric polynomial fit may be almost 50, making much worse accuracy of the approximation.

The NAV algorithm is effective for all three types of eclipsing variables[44]. In a case of multi-color observations, and additional statistical dependencies, phenomenological modeling allows estimating of physical parameters[45]. In the case of observations in one filter only, the approximation in a physical model may be of almost the same quality not for a single set of parameters, but for some region in the parameter space[46].

**Appoximations in separate intervals near specific points**

In some signals, there are parts of constant values, which are interrupted by intervals of activity. These may be transitions between some levels, outbursts or eclipses. In these cases, there is no need for all-time monitoring, and the observations are planned either after the "outburst alert" by someone observing many objects rarely (one of many objects will be at an outburst with a high probability), or before and after a predicted time of event (minimum or maximum). Even for "almost periodic" signals, there may be observed additional mechanisms of variability, e.g. "O-C" variability in eclipsing variables – due to a presence of a third body, apsidal motion, mass and angular momentum transfer. There are special observational programs to "catch the minimum", and the corresponding tables are published in papers and on-line catalogues. A large compilation of such times of minima (ToM) was recently published, e.g. by D. Tvardovskyi[47], and analyzed, with a special interest to stellar-mass third bodies at elliptic orbits[48].

A set of functions for determination of the times of minima with the best accuracy was reviewed[49]. The accuracy may be significantly different for the same data and same number of parameters. Special shapes are important, but they are applied separately to the short data, thus decreasing the number of parameters.

The classical approach is to use polynomials with a degree, which provides the best accuracy of ToM. As an example, there is a compilation of 6509 ToM for 147 semi-regular stars[50]. Another method was a spline with alternating degrees 1-2-1, i.e. the method of "asymptotic parabola"[51,52]. For more extended intervals up to "from previous to next extremum", the approximation may be changed to 3-interval cubic spline[25]. For complete eclipses, the approximating functions were compared[53]. For shorter intervals, covering the bottom of eclipse, and only a part of the ascending/descending branches, the "wall-supported" functions were proposed[54].

For determination of the moment $t_L$ of crossing of some limiting value $x_L$, one may either solve the equation $x_C(t_L) = x_L$, or to make an inverse approximation $t_C(x)$ and then compute $t_C(x_L)$.

The simplest approximation – the line – was used for determination of moments in a case of very abrupt changes of brightness in the object V808 Aur[55].

---

## MULTI-COMPONENT VARIABILITY OF PULSATING STARS

A review on semi-regular pulsating variable stars was published recently[56].

For other group of pulsating variable stars with multi-component variability – RV Tau-type stars – the classification based on complexity of the periodograms is discussed. Periodograms are previously classified by their shape into three groups according to the presence or absence of certain structures: two peaks in a 2:1 ratio, the presence of satellites of these peaks indicating the result of beats[57].

For the large-amplitude pulsating variables, in an addition to approximation using trigonometric polynomial of statistically optimal order $s$, an atlas of the phase plane diagrams "$dx_C(\phi)/d\phi$ vs $x_C(\phi)$" was compiled and analyzed[58,59].

The relation between optical variability and maser emission was studied in the articles[60,61,62]

## MAGNETIC CATACLYSMIC VARIABLES

Cataclysmic variables are close binary systems usually consisting of a white dwarf and a red dwarf filling the Roche lobe. Depending in the degree of influence of the magnetic field of the white dwarf onto accretion, the rotation may be synchronous (AM Her-type), "idling" (BY Cam-type) or "fast" (DQ Her-type).

We regularly obtain photometric data within the collaboration with Observatory and Planetarium of M. R. Stefanik in Hlohovec and Vihorlat Astronomical Observatory (Slovakia), Fort Skala Astronomical Observatory of the Jagiellonian University (Poland) and other institutions worldwide, including time series from international databases.

The rotational evolution of white dwarf is analyzed using the O-C diagrams. For determination of the moments of minima of orbital variability and (simultaneously) maxima of the spin variability, the two-period model is used, which implemented in the software MCV.

Complicated variations of different types were found in V405 Aur[63] in FO Aqr[64]. Fast spin-down was detected in V2306 Cyg[65], MU Cam[66] and EX Hya[67].

In MU Cam, we have investigated the periodic modulation of the spin phases with the orbital phase[68]. As a possible source of the unexpected scatter on this figure we have investigated the dependency of spin maxima timings on orbital phase described by Kim et al.[69].

The preliminary value of the period[70] of the intermediate polar V1323 Aql was improved[66].

The idling of the magnetic white dwarf in the asynchronous polar BY Cam was studied during the international observational campaign "Noah-2"[71].

## CONCLUSIONS

In this review, we present some advanced algorithms and programs used in our scientific school with short description of types of astrophysical systems, which we study. These are variable stars of different types, with a special attention to complicated systems, which exhibit various mechanisms of variability: pulsating variable stars, eclipsing, cataclysmic and symbiotic binary systems.

As an example of relations to engineering, one may refer to vibrations, stability of mechanisms[72,73,74]. Many mathematical equations are common in Science, Technics and Humanities.

The majority of observational results have been obtained in a close collaboration with astronomers in the Universities and astronomical observatories in Odessa, as well as in Korea, Slovakia, Poland, USA, Greece and other countries.

---

[64] Breus V.V., Andronov I.L., Hegedus T., Dubovsky pp.A., Kudzej I. (2012). – Two-period variability of the intermediate polar FO Aqr. – *Advances in Astronomy and Space Physics*, Vol. 2, pp.9-10. ADS: 2012AASP....2....9B

[65] Breus V., Petrík K., Zola S. (2019). Detection of white dwarf spin period variability in the intermediate polar V2306 Cygni. – *Monthly Notices of the Royal Astronomical Society,* Volume 488, Issue 4, pp.4526-4529. DOI: https://doi.org/10.1093/mnras/stz2062 . ADS: 2019MNRAS.488.4526B

[66] Petrik K., Breus V.V., Andronov I.L., et al. (2015). Spin Period Study of the Intermediate Polars MU Cam, V2306 Cyg and V1323 Her. – *Astronomical Society of the Pacific Conf. Ser.*, Vol. 496, pp.252-253. ADS: 2015ASPC..496..252P

[67] Andronov I.L., Breus V.V. (2013). Variability of the Rotation Period of the White Dwarf in the Magnetic Cataclysmic Binary System EX Hya. – *Astrophysics*, Vol. 56, no. 4, pp.518-530. – DOI: https://doi.org/10.1007/s10511-013-9304-7 . ADS: 2013Ap.....56..518A

[68] Parimucha Š., Dubovsky pp.A., Kudzej I., Breus V., Petrik K. (2020). About the dependency of the spin maxima on orbital phase in the intermediate polar MU Cam. - *Contributions of the Astronomical Observatory Skalnaté Pleso*, vol. 50, no. 2, pp. 618-620. DOI: https://doi.org/10.31577/caosp.2020.50.2.618 . ADS: 2020CoSka..50..618P

[69] Kim Yong-Gi, Andronov I.L., Park Sung-Su, Chinarova L.L., Baklanov A.V., Jeon, Young-Beom (2005). Two-Color VR CCD Photometry of the Intermediate Polar 1RXS J062518.2+733433. - *Journal of Astronomy and Space Sciences*. Vol. 22, no. 3, pp. 197-210. DOI: https://doi.org/10.5140/JASS.2005.22.3.197 . ADS: 2005JASS...22..197K

[70] Andronov I.L., Kim Yonggi Kim, Yoon Joh-Na, et al. (2011). Two-Color CCD Photometry of the Intermediate Polar 1RXS J180340.0+401214. - *Journal of the Korean Astronomical Society*, vol. 44, no. 3, pp. 89-96. DOI: https://doi.org/10.5303/JKAS.2011.44.3.089 . ADS: 2011JKAS...44...89K

[71] Andronov I.L., Antoniuk K.A., Breus V.V., et al. (2008). Idling magnetic white dwarf in the synchronizing polar BY Cam. The Noah-2 project. - *Central European Journal of Physics*, Volume 6, Issue 3, pp.385-401. DOI: https://10.2478/s11534-008-0076-3 ADS: 2008CEJPh...6..385A

[72] Bol'shakov V.I., Butsukin V.V. (2005). Osobennosti opredeleniya parametrov bieniy pri issledovanii dinamiki mashin [Features of determining the parameters of the beats in the study of the dynamics of machines]. *Fundamental'nye i prikladnye problemy chernoy metallurgii*: Sb. nauchn. tr., Dnipropetrovs'k.: IChM NAN Ukraïni, no. 10, pp. 300-306.

[73] Konoplev A.V., Kononova O.N. (2018). Otsenka soprotivleniya ustalosti vosstanovlennykh detaley sudovykh mashin i mekhanizmov [Assessment of fatigue resistance of reconditioned parts of ship engines and mechanisms]. *Visnik Odes'kogo natsional'nogo mors'kogo universitetu*, no.2(55), pp. 68-74.

[74] Barenblatt G.I. (2003). *Scaling*. Cambridge University Press.

However, we discuss mainly mathematical methods, which may be applied to analysis of signal of any nature – in computer science, engineering, economics, social studies, decision making etc. A variety of types of signals need a diversity of adequate complementary specific methods, in an addition to common algorithms.

**Information about the authors:**
**Andronov I.L.,**
Doctor (Dr. hab.) of Physical and Mathematical Sciences, Professor,
Academician of the "Academy of the High School of Ukraine",
Chair of the Department "Mathematics, Physics and Astronomy",
Odessa National Maritime University
34, Mechnikova str., Odessa, 65029 Ukraine
ORCID ID: orcid.org/ 0000-0001-5874-0632

**Breus V.V.,**
PhD in Physical and Mathematical Sciences,
Associate Professor of the Department "Mathematics, Physics and Astronomy",
Odessa National Maritime University
34, Mechnikova str., Odessa, 65029 Ukraine
ORCID ID: orcid.org/ 0000-0002-7535-2241

**Kudashkina L.S.,**
PhD in Physical and Mathematical Sciences,
Associate Professor of the Department "Mathematics, Physics and Astronomy",
Odessa National Maritime University
34, Mechnikova str., Odessa, 65029 Ukraine
ORCID ID: orcid.org/ 0000-0002-8482-9240